\documentstyle[psfig]{l-aa}

\begin{document}

\thesaurus{02 (12.07.1; 12.12.1; 12.03.4)} 

\title{Weighing neutrinos: weak lensing approach}

\author{Asantha R. Cooray}

\institute{Department of Astronomy and Astrophysics, University of
Chicago, Chicago IL 60637, USA. E-mail: asante@hyde.uchicago.edu}

\date{Received: 15 March 1999 ; accepted: 12 May 1999}
\maketitle

\begin{abstract}

We study the possibility for a measurement of neutrino mass using
weak gravitational lensing. The presence of non-zero mass
neutrinos leads to a suppression of power at small scales
and reduces the expected weak lensing signal.
The measurement of such a suppression in the weak
lensing power spectrum allows a direct measurement of the neutrino
mass, in contrast to various other experiments which only allow 
mass splittings between two neutrino species.
Making reasonable assumptions on the accuracy of cosmological
parameters, we suggest that a weak lensing survey of
100 sqr. degrees can be  easily used to detect neutrinos  down to
a mass limit of $\sim$ 3.5 eV at the 2$\sigma$ level. This limit is lower
than current limits on neutrino mass, such as from the Ly$\alpha$
forest and SN1987A. 
An ultimate weak lensing survey of $\pi$ steradians down to a
magnitude limit of 25 can be used to detect neutrinos down to
a mass limit of 0.4 eV at the 2$\sigma$ level, provided
that other cosmological parameters will be known to an accuracy
expected from cosmic microwave background spectrum using
the MAP satellite. With improved parameters estimated from
the PLANCK satellite, the limit on neutrino mass from weak lensing
can be further lowered by another
 factor of 3 to 4. For much smaller surveys ($\sim$ 10 sqr. degrees)
that are likely to be first available in the near future 
with several wide-field cameras,  the presence of neutrinos can 
be safely ignored when deriving conventional cosmological parameters
such as the mass density of the Universe.
However, armed with cosmological parameter
estimates with other techniques,  even such small area  surveys
allow a strong possibility to investigate the presence 
of  non-zero mass neutrinos.

\end{abstract}

\keywords{cosmology: theory --- dark Matter ---
large-scale structure of Universe ---- gravitational lensing}

\section{Introduction}

It is now well known that the statistical analysis of weak lensing
effects on background galaxies due to foreground large scale
structure can be used as a probe of cosmological parameters,
such as the matter density and the cosmological constant, and the
projected mass distribution of the Universe (e.g., 
Bernadeau et al. 1997; Blandford et al. 1991; 
Jain \& Seljak 1997; Kaiser 1998;
Miralda-Escud\'e 1991;  Schneider et al. 1998).
Given that weak gravitational lensing results from the projected mass
distribution, the statistical properties of weak lensing,
such as the two point function of the shear distribution, reflects certain
aspects associated with the projected matter distribution.
With growing interest in weak gravitational lensing
surveys, several studies have now explored the accuracy to which
conventional cosmological parameters, such as the
mass density of the Universe and the cosmological constant
 can be determined (e.g., Bartelmann \& Schneider 1999;
van Waerbeke et al. 1999).

Beyond primary cosmological parameters, such as the mass density,
the projected matter distribution is also affected by the presence of
neutrinos with non-zero masses. For example, 
when non-zero mass neutrinos are present, a strong suppression of power
in the mass distribution occurs at scales below the 
time-dependent free streaming scale (see, e.g., Hu \& Eisenstein 1998).
The detection of such suppression, say in the density power spectrum,
allows a direct  measurement of the neutrino mass, in contrast to
various particle  physics based neutrino experiments which only allow 
measurements of mass differences, or splittings, between different
neutrino species (e.g., Super Kamiokande experiment; Fukuda et al. 1998). 
The direct astrophysical probes of neutrino masses include
time-of-flight from a core-collapse supernova (e.g., Totani 1998;
Beacom \& Vogel 1998; see, Beacom 1999 for a review) and large scale
structure power spectrum. The suppression of power at small scales due to
neutrinos can easily  
be investigated with the galaxy power spectrum from wide-field
redshift surveys, such as the Sloan Digital Sky Survey 
(SDSS\footnote{http://www.sdss.org/}; e.g., Hu et al. 1997), however,
such measurements are subjected to unknown biases between
galaxy and matter distribution and its evolution with redshift.
Therefore, an understanding of bias and its evolution may first
be necessary before making a reliable measurement of neutrino
mass using galaxy power spectra. On the contrary, a measurement
of the power spectrum unaffected by such effects allows
a strong possibility to measure the neutrino mass.
Such a possibility should soon be available with weak gravitational
lensing surveys through the
measured weak lensing power spectrum directly, which probes the matter
power spectrum through a convolution of the redshift distribution
of sources and distances.
Thus, it is expected that the weak lensing power 
spectrum can also allow a determination of the suppression due to
neutrinos, and thus, a direct measurement of the  neutrino mass. 

In addition to neutrino mass, weak lensing also allows determination
of several cosmological parameters, including
matter density ($\Omega_m$) and the cosmological constant 
($\Omega_\Lambda$). However, there are
large number of cosmological 
probes that essentially measure these parameters. For
example, luminosity distance measurements to Type Ia supernovae
at high redshifts and gravitational lensing statistics
allow the determination of
$\Omega_m$ and $\Omega_\Lambda$ (e.g., Cooray et al. 1998;
Perlmutter et al. 1998; Riess et al. 1998), 
or rather $\Omega_m-\Omega_\Lambda$,
while the location of the first Doppler peak in the cosmic microwave
background (CMB) power spectrum allows a determination of these two
quantities in the orthogonal direction ($\Omega_m+\Omega_\Lambda$;
White 1998). 
When combined (e.g., Lineweaver 1998), 
these two parameters can be known to a high accuracy
reaching a level of $\sim$ 5\%, when the expected measurements on
SNe and CMB experiments over the next decade, such as MAP, are 
considered (e.g.,
Tegmark et al. 1998). Since these measurements are not 
sensitive to neutrinos, therefore, it is necessary that one returns to
probes which are strongly sensitive to neutrinos to obtain important
cosmological information on their presence. This is the primary
motivation of this paper: weak lensing is highly suitable
for a neutrino mass measurement when compared to various other probes
of cosmology, including CMB.
The direct measurement of galaxy power spectrum
also allows a measurement of neutrino mass, as has been discussed in
Hu et al. (1997), however as discussed
above, such a measurement can be contaminated by bias
and its evolution. 

The other motivation for this paper comes through
the cosmological importance of neutrinos (see, Ma 1999 for a recent
review on this subject).  Current upper bounds on neutrino masses
range from 23 eV based on SN 1987A neutrino arrival time delays
to 4.4 eV using recent oscillation experiments, assuming that
3 degenerate neutrino species are present (Vernon Barger,
priv. communication; Fogli et al. 1997. Recently, Croft et al.
(1999) determined that $m_\nu < 5.5$ eV at the 95\% level using
the Ly$\alpha$ forest for all $\Omega_m$ values and
$m_\nu < 2.4 (\Omega_m/0.17 - 1)$ eV for $0.2 \la \Omega_m \la 0.5$
(95\% confidence).  A rather definite upper limit on
neutrino mass is 94 eV, which is the mass limit to
produce a normalized cosmological mass density of 1, while
according to Ma (1999), a rather conservative 
cosmological limit on neutrino
mass presently is $\sim$ 5 eV.
However, apart from these limits, several
studies still suggest the possibility that the neutrino
mass can be as high as 15 eV (e.g., Shi \& Fuller 1999), therefore,
it is safe to say that neutrino mass or limit on its mass is not
strongly constrained. The neutrino mass is one of the important
cosmological parameters, and thus, it is necessary that suitable
probes which allow this measurement, beyond mass splitting
measurements allowed by particle physics experiments,
be studied.

In this paper, we explore the possibility for a neutrino
mass measurement with weak lensing surveys and suggest that weak lensing
can be used  as a strong probe of the neutrino mass, provided
that one has adequate knowledge on 
the uncertainties of basic cosmological parameters  will other
techniques. In a recent paper, Hu \& Tegmark (1999)
explored the full parameter space of wide-field weak lensing surveys combined
with future cosmic microwave background (CMB) satellites.
A recent review on weak lensing could be found in Mellier (1998).
In Sect.~2, we discuss the effect of neutrinos in the
weak lensing convergence power spectrum and calculate accuracies
to which the neutrino mass can be determined.
We follow the conventions that the Hubble constant,
$H_0$, is 100\,$h$\ km~s$^{-1}$~Mpc$^{-1}$ and
 $\Omega_i$ is the fraction of the critical
density contributed by the $i$th energy component: $b$ baryons, $\nu$
neutrinos, $m$ all matter species (including baryons and neutrinos)
and $\Lambda$ cosmological constant.

\section{Weak Lensing Power Spectrum}

\subsection{Effective Convergence Power Spectrum}

Following Kaiser (1998) and Jain \& Seljak (1997), we can write the
power spectrum of convergence due to weak gravitational lensing as:
\begin{equation}
P_\kappa(l) = l^4 \int d\chi \frac{g^2(\chi)}{r^6(\chi)}
P_\Phi\left(\frac{l}{r(\chi)},\chi\right),
\end{equation}
where $\chi$ is the radial comoving distance related to redshift
$z$ through:
\begin{equation}
\chi(z) =  \frac{c}{H_0}\int_{0}^{z} dz' \left[ \Omega_m (1+z')^3 +
\Omega_k (1+z')^2 + \Omega_\Lambda \right]^{-1/2},
\end{equation}
and $r(\chi)$ is the comoving angular diameter distance
written as $r(\chi) = 1/\sqrt{-K} \sin \sqrt{-K}\chi,\chi,
1/\sqrt{K}\sinh \sqrt{K}\chi$ for closed, flat and open models
respectively with $K = (1-\Omega_{\rm tot})H_0^2/c^2$.
In Eq.~1, $P_\Phi(k=\frac{l}{r(\chi)},\chi)$ is the time-dependent
three dimensional power spectrum of the Newtonian potential
which is related to the density power spectrum, $P_\delta(k)$, through the
Poisson's equation (e.g., Eq.~2.6 of Schneider et al. 1998)
and $g(\chi)$ weights the background source distribution by the
lensing probability:
\begin{equation}
g(\chi) = r(\chi) \int_{\chi}^{\chi_H}
\frac{r(\chi'-\chi)}{r(\chi')} W_\chi(\chi')
d\chi'.
\end{equation}
Here, $\chi_H$ is the comoving distance to the horizon.

Following Kaiser (1998) and Hu \& Tegmark (1999), 
we can write the expected uncertainties in the weak lensing
convergence power spectrum as:
\begin{equation}
\sigma(P_\kappa)(l) = \sqrt{\frac{2}{(2\l+1)f_{\rm sky}}}\left(
P_\kappa(l) + \frac{\langle \gamma^2 \rangle}{n_{\rm mag}}\right),
\end{equation} 
where $f_{\rm sky}$ is the fraction of the sky covered by a survey,
$\sqrt{\langle \gamma^2 \rangle} \sim 0.4$ is the intrinsic non-zero
ellipticity of background galaxies and $n_{\rm mag}$ 
is the surface density of galaxies down to the magnitude limit
of the survey. Thus, Eq.~5, accounts for three sources of noise in the
weak lensing power spectrum: cosmic variance, shot-noise in the
ellipticity measurements and the number of galaxies available to make such
measurements from which the weak lensing properties are
derived (see, e.g., Schneider et al. 1998 and Kaiser 1998 for further details).
We take $n_{\rm mag}$ to be $6.5 \times 10^{8}$ sr$^{-1}$ 
down to R magnitude of 25 and $4 \times 10^{9}$ sr$^{-1}$ down to
R of 27, which were determined based on galaxy number counts of deep
surveys such  as the Hubble Deep Field.

Following Schneider et al. (1998), we parameterize the source
distribution, $W_\chi(\chi)$, as a function of redshift, 
$W_z(z)$: 
\begin{equation}
W_z(z) =
\frac{\beta}{\Gamma\left[\frac{1+\alpha}{\beta}\right] z_0} \left(\frac{z}{z_0}\right)^\alpha
\exp\left[-\left(\frac{z}{z_0}\right)^\beta\right].
\end{equation}
Such a distribution has been observed to provide a good fit to the
observed redshift distribution of galaxies (e.g., Smail et al. 1995).

\subsection{Linear and Nonlinear Power Spectra}

Since we are considering non-zero mass neutrinos, it is necessary that
both a linear and a nonlinear power spectrum which takes in
to account for such neutrinos be considered. In order to obtain the
linear power spectrum, we follow Hu \&
Eisenstein (1998) and Eisenstein \& Hu (1999) 
and consider the MDM (mixed dark matter) transfer
and growth functions appropriate for massive neutrinos
as well as baryons. We use fitting formulae presented therein which agree
with numerical calculations at a level of 1\%. Both neutrinos
and baryons affect the standard power spectrum by 
suppressing power at small scales below the free-streaming length.
The small scale suppression due to neutrinos can be written as:
\begin{equation}
\left(\frac{\Delta P}{P}\right) \sim -8\frac{\Omega_\nu}{\Omega_m}
\sim
-0.8\left(\frac{m_\nu}{{\rm 1\,eV}}\right)\left(\frac{0.1N}{\Omega_m
h^2}\right),
\end{equation}
where $N$ is the number of degenerate neutrinos. 
Assuming the standard model for neutrinos with a temperature $(4/11)^{1/3}$
that of the CMB, we can write $\Omega_\nu$ based on neutrino
mass, $m_\nu$ (in eV), and the number of degenerate neutrino species,
$N$, as $\Omega_\nu = N(m_\nu/94)h^{-2}$. We assume integer number of
neutrino species that can amount up to three. 
The suppression of
power is proportional to the ratio of hot matter density of neutrinos
to cold matter density; in low $\Omega_m$ cosmological models,
currently preferred by observations, the suppression of power is 
much larger than in an Einstein-de Sitter universe with the same
amount of neutrinos. In fact, for low $\Omega_m$ models, massive
neutrinos of mass $\sim$ 1 eV, contribute to a 100\% suppression of
power compared with no neutrinos.

In addition to the linear power spectrum,
given the time-dependence, it is necessary  that the non-linear
evolution of the density power spectrum be fully taken into account
when calculating the convergence power spectrum given in Eq.~1. The
importance of the non linear evolution of the power spectrum on weak
lensing statistics was first discussed in Jain \& Seljak (1997)
for standard $\Lambda$CDM cosmological models involving
$\Omega_m$ and $\Omega_\Lambda$.  
There are several approaches to obtain the nonlinear evolution,
however, for analytical calculations, fitting functions
are strongly preferred over detailed numerical work.
In Peacock \& Dodds (1996), the evolved density power spectrum
was related to the linear power spectrum through a function
$F(x)$, where $x$ was calibrated against numerical simulations
in standard CDM models. According to Peacock \& Dodds (1996), the
nonlinear power spectrum $P_\delta$ is related to the linear power
spectrum, $P_\delta^L(k_L)$, through: $k^3P_\delta(k)/(2\pi^2) =
F\left[k_L^3P_\delta^L(k_L)/(2\pi^2)\right]$ where $k_L =
\left[1+ k^3P_\delta(k)/(2\pi^2)\right]^{-1/3}k$. 
We refer the reader to Peacock \& Dodds (1996) for the functional
form of $F(x)$. 

Since we are now allowing for the presence of
massive neutrinos, as well as baryons,
it is necessary that we consider whether the fitting
function given in Peacock \& Dodds (1996) is reliable for the
present calculation as these two species were not included in their
simulations; Smith et al. (1998) compared the
Peacock \& Dodds (1996) formulation  against
 MDM numerical simulations and suggested a possible agreement
between the two when spectral index for Peacock \& Dodds (1996)
fitting formula was calculated using MDM power spectrum.
However, recently, Ma (1998) suggested that this agreement was
only due to poor resolution of numerical simulations used in
Smith et al. (1998). According to Ma (1998), Peacock \& Dodds
(1996) formulation disagrees with numerical data at the level of
10\% to 50\%. Therefore, instead of the Peacock \& Dodds (1996)
approach, we use the fitting function given in Ma (1998) in the present
calculation which was now shown to agree with numerical simulations at a
level of 3\% to 10\% for $k \la 10\; h$ Mpc$^{-1}$ 
out to a redshift of $\sim 4$. For higher scales and redshifts,
 agreement is only reached at a level of 15\% against
numerical simulations. For the present calculation
involving a redshift distribution that peaks at redshifts lower
than 4 with scales of interest lower than 10 $h^{-1}$ Mpc$^{-1}$, 
the fitted formulation is reasonably adequate. 
Since this fitting formulae is still not widely used, 
compared to Peacock \& Dodds (1996)
formulae, we reproduce them here for interested readers.
The nonlinear power spectrum is related to the linear power spectrum
through (Ma 1998):
\begin{eqnarray}
       && {\Delta(k)\over \Delta_L(k_L)} = G\left({\Delta_L(k_L) \over
        g_0^{1.5}\,\sigma_8^\beta} \right) \,,\nonumber\\ &&
        G(x)=[1+\ln(1+0.5\,x)]\,{1+0.02\,x^4 + c_1\,x^8/g^3 \over
        1+c_2\,x^{7.5}}\,,
\end{eqnarray}
where $\Delta(k)\equiv k^3P_\delta(k)/(2\pi^2)$ is the density variance in the
linear and nonlinear regimes\footnote{The notations used
by Peacock \& Dodds (1996) and Ma (1998) differs in that $P(k)$ defined
in Ma (1998) refers to $P(k)/(2\pi)^3$ in Peacock \& Dodds (1996).}. 
Similar to Peacock \& Dodds (1996),
the nonlinear scale is related to the linear scale through:
\begin{equation}
k_L =\left[1+ \Delta(k)\right]^{-1/3}k.
\end{equation}
Instead of the effective spectral index $n_{\rm eff}$ used in
Peacock \& Dodds (1996), the formalism uses $\sigma_8$ 
which is the rms linear mass fluctuation on $8\,h^{-1}$ Mpc scale
evaluated at the redshift of interest. 
The numerical simulations suggest that $n_{\rm eff}+3$ is related to
$\sigma_8$ through $n_{\rm eff}+3 \sim \sigma_8^\beta$ where 
$\beta=0.7+10\Omega_\nu^2$.  The functions $g_0=g(\Omega_m,\Omega_\Lambda)$ and
$g=g(\Omega_m(z),\Omega_\Lambda(z))$ are, 
respectively, the relative growth factor for
the linear density field evaluated at present
 and at redshift $z$, for a model with a
present-day matter density $\Omega_m$ and a cosmological constant 
$\Omega_\Lambda$.  A fitting formula for
$g(\Omega_m,\Omega_\Lambda)$ is (Carroll et al. 1992):
\begin{eqnarray}
&&  g ={5\over 2}\Omega_m(z) \, \\ && [
      \Omega_m(z)^{4/7}-\Omega_\Lambda(z)+\left(1+ \Omega_m(z)/2\right) \left(1+ \Omega_\Lambda(z)/70
      \right)]^{-1}\, \nonumber .
\end{eqnarray}
According to Ma (1998), for CDM and LCDM models, a
good fit is given by $c_1=1.08\times 10^{-4}$ and $c_2=2.10\times
10^{-5}$, while $c_1=3.16\times 10^{-3}$
and $c_2=3.49\times 10^{-4}$ for MDM models with $\Omega_\nu$
of $\sim$ 0.1 and $c_1 =6.96\times 10^{-3}$
and $c_2=4.39\times 10^{-4}$ for MDM models with $\Omega_\nu$
of $\sim$ 0.2. For all other $\Omega_\nu$ values, usually less than
0.1 for neutrino masses of current interest, we interpolate between
the published values of $c_1$ and $c_2$ by Ma (1998). This procedure
should be approximate, but for higher precision, numerical simulations
would be required to determine $c_1$ and $c_2$ at individual $\Omega_\nu$
values. In general, the weak lensing convergence power spectrum depends on six
cosmological parameters: $\Omega_m$, $\Omega_\Lambda$, $\Omega_b$, 
$\Omega_\nu$,  $n_s$ the primordial scalar tilt and $\delta_H$
the normalization of the density power spectrum. 
Also, throughout this
paper, we take a flat model in which $\Omega_m+\Omega_\Lambda=1$.
Such a cosmology is motivated by both inflationary scenarios
 and current observational data. 
For $\delta_H$ we use COBE normalizations as presented by
Bunn \& White (1997) and also consider galaxy cluster based
normalizations, $\sigma_8$, from Viana \& Liddle (1998).

\begin{figure*}
\centerline{\psfig{file=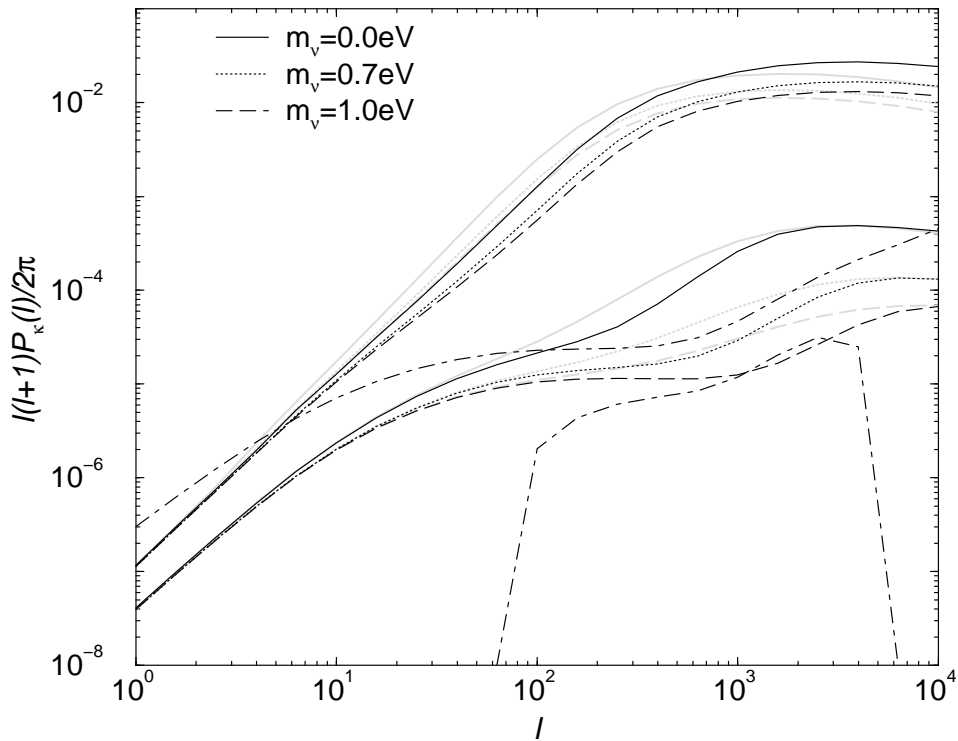,width=5.0in,angle=-90}}
\caption{Weak Lensing Power Spectrum for two COBE normalized models
involving non-zero mass neutrinos. The upper set of curves correspond
to a flat cosmological model involving $\Omega_m=0.75$,
$\Omega_b=0.05$, $h=0.65$, $n_s=1$, while the lower set of curves are
for $\Omega_m=0.35$ with other parameters as above. In {\it grey},
we show weak lensing power spectra using Peacock \& Dodds
(1996) fitting function for nonlinear evolution of the power spectrum,
calculated assuming its validity for MDM cosmologies, while
dark lines show weak lensing power spectra 
for recently updated fitting functions for MDM 
cosmologies by Ma (1998).  In {\it dot-dashed} lines, we show
expected errors in the power spectrum measurement from
a weak lensing survey of 25 $\times$ 25 deg$^{2}$ down to the
magnitude limits of 25 in R. Such a survey is expected to be available
in near future with wide-field cameras such as MEGACAM (Boulade et al. 1998).}
\end{figure*}

In Fig.~1, we show two sets of COBE normalized weak lensing power spectra
considering the presence of non-zero mass neutrinos. The upper and
lower curves represent two cosmological models with high and low 
$\Omega_m$ values and computed assuming the redshift of
background sources as given in Eq.~4, with $\beta=1.5$ and $\alpha=2.0$. 
As shown, non-zero mass neutrinos suppress power
at large $l$ values, and this effect is significant for low
$\Omega_m$ models. This is primarily due to that fact that the
suppression of power is directly proportional to the ratio of
$\Omega_\nu/\Omega_m$. In addition, we have also
shown the expected 1$\sigma$ uncertainty in the power spectrum
measurement for a survey of size 625 deg$^{2}$ down to magnitude
limit of 25 in R. It is likely that weak lensing surveys down
to R of 25 within an area of 100 deg$^2$ will be available in
the near future, and that
the area coverage  would steadily grow as high as several thousand
square degrees over the next decade. 
As shown in Fig.~1, reliable measurements of the power spectrum
is likely when $l$ is between 100 and 3000. This is the same range in
which neutrinos suppress power. Such effects do not exist, for
example, in the CMB anisotropy power spectrum;
low redshift probes of the matter power spectrum provide ideal ways to weigh
neutrinos.

\subsection{Cosmic Confusion?}

However, there are alternative possibilities which can mimic
neutrinos. In Fig.~2, as examples, we illustrate two possibilities
which can produce a similar power spectrum as a model
involving $\Omega_m$ of 0.35 and $m_\nu$ of 0.7 eV;
When $m_\nu$ is 1 eV, increasing the primordial scalar tilt by 30\% can
mimic the original power spectrum, while in a model with zero mass
neutrinos, increasing the baryon content by 80\% can produce essentially
the original power spectrum. Such effects are essentially what can be 
described as {\it cosmic confusion}, and thus, careful measurements of
cosmological parameters are needed to weigh neutrinos even with weak
lensing.

\begin{figure}
\centerline{\psfig{file=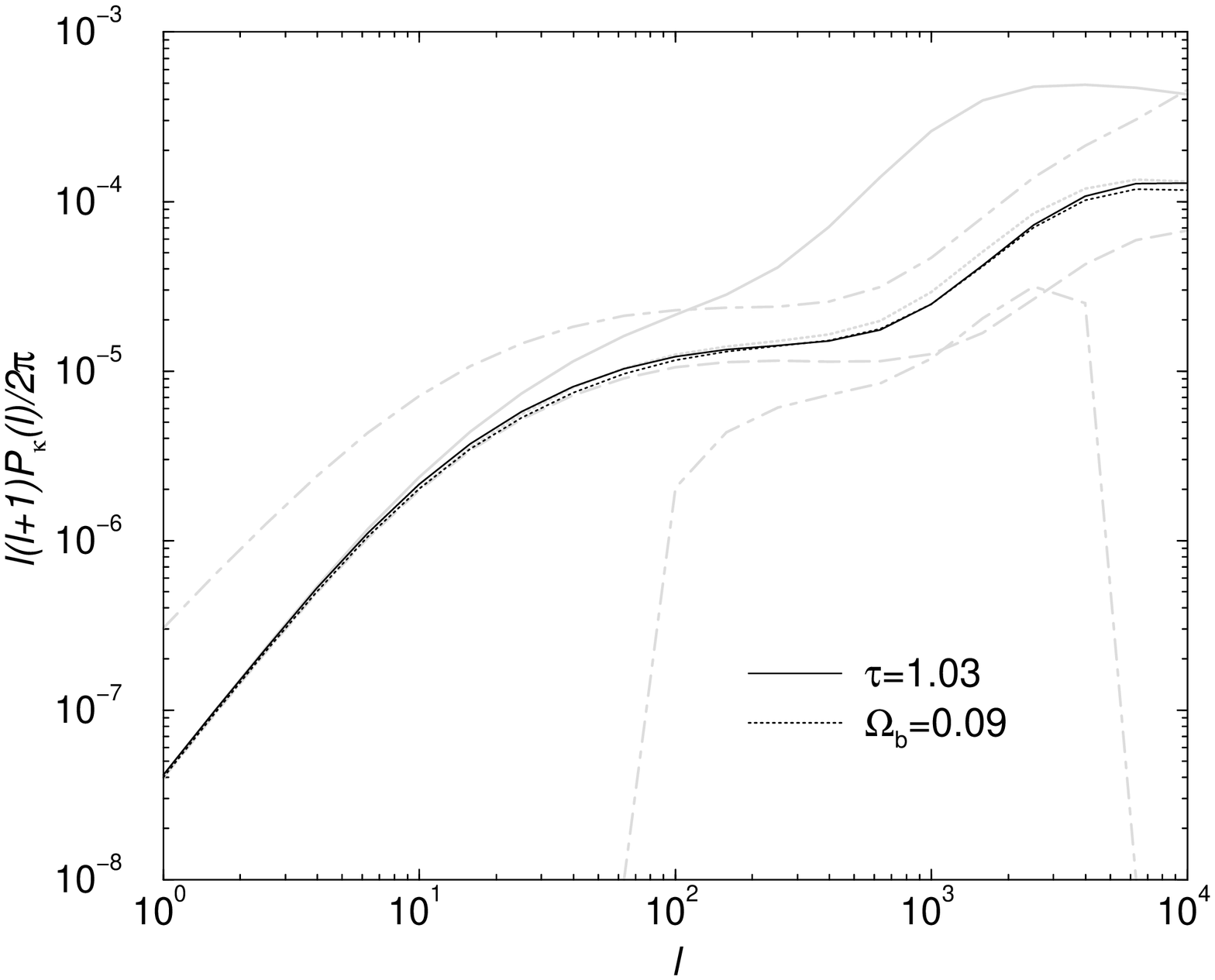,width=3.6in,angle=-90}}
\caption{Cosmic confusion: Two alternate models involving
changes in the baryon content or the scalar tilt produce essentially the
same power spectrum as a model involving 0.7 eV neutrinos. The {\it grey}
curves are same as Fig.~1.}
\end{figure}

\section{Neutrino Mass Measurement}

In order to investigate the possibility for a neutrino mass
measurement, we consider the so-called Fisher information matrix
(e.g., Tegmark et al. 1997) with six cosmological parameters 
 that define the weak lensing
power spectrum. The Fisher matrix $F$ can be written as:
\begin{equation}
F_{ij} = -\left< \partial^2 \ln L \over \partial p_i \partial p_j
 \right>_{\bf x},
\end{equation}
where $L$ is the likelihood of observing data set ${\bf x}$
given the parameters $p_1 \ldots p_n$.
Following the Cram\'er-Rao inequality, 
no unbiased method can measure the {\it i}th parameter 
with standard deviation  less than $(F_{ii})^{-1/2}$ if other parameters
are known, and less than $[(F^{-1})_{ii}]^{1/2}$ if other parameters
 are estimated from the  data as well. Since  Eq.~6 is usually
calculated  assuming a prior cosmological model, 
the estimated errors on the parameters of this
underlying model can be dependent on prior assumptions.

Assuming a Gaussian and uncorrelated distribution for uncertainties, one
can easily derive the Fisher matrix for weak lensing as\footnote
{Note the minor correction to Eq.~4 of Hu \& Tegmark (1999)}:
\begin{equation}
F_{ij} = \sum_{l=l_{\rm min}}^{l_{\rm max}} \frac{f_{\rm sky} (2l
+1)}{2 \left(P_\kappa(l)+\frac{\langle \gamma^2 \rangle}{n_{\rm mag}}\right)^2}
\frac{\partial P_\kappa(l)}{\partial p_i}
\frac{\partial P_\kappa(l)}{\partial p_j}.
\end{equation}

As illustrated in Fig.~2, in order to make a reliable measurement
of neutrino mass, it is necessary that one consider external
measurements of cosmological parameters. Such measurements can come from
variety of probes such as Type Ia supernovae, galaxy clusters, CMB,
gravitational lensing etc. Here, we take both a conservative approach
with large uncertainties for the cosmological parameters based on
other techniques and a more optimistic approach motivated by the
expected uncertainties from future surveys.
In our conservative model, we use following errors:
$\sigma(\Omega_m)=0.2$, $\sigma(\Omega_b) = 0.1 \Omega_b$,
$\sigma(h) =0.2$, $\sigma(n_s)=0.1$, $\sigma(\ln \delta_H) = 0.5$,
while in our optimistic model, we use 
$\sigma(\Omega_m)=0.07$, $\sigma(\Omega_b) = 0.0025h^{-2}$,
$\sigma(h) =0.1$, $\sigma(n_s)=0.06$, $\sigma(\ln \delta_H) = 0.3$.
These errors are in fact worser than what is expected to be measured
from PLANCK\footnote{http://astro.estec.esa.nl/Planck/; also, ESA document D/SCI(96)3.}, 
but is similar to what could be achieved with
a mission such as MAP\footnote{http://map.gsfc.nasa.gov/}. We consider
a fiducial model in which 
$\Omega_b=0.019$, consistent with current observations, $h=0.65$ and 
$n_s=1.0$ and
normalization  based on COBE. In addition, we also consider an
alternative normalization to the power spectrum based on measurements
of  $\sigma_8$  ($=0.56\Omega_m^{-0.47}$) following 
Viana \& Liddle (1998). We also consider variations to the above
fudicial model and marginalize over the
uncertainties to obtain the 2$\sigma$ detection limit of neutrinos for
various weak lensing surveys. We only use the information on
the power spectrum between $l$ values of 100 and 5000. At $l$ values
below 100, cosmic variance dominate the measurement while
at $l > 5000$ the finite number of galaxies and their ellipticies 
contribute to the increase in power spectrum measurement uncertainties.

In Fig.~3, we summarize our results: solid lines show the
expected 2$\sigma$ detection limit  for our conservative errors
while dashed lines show the detection limits for more optimistic
errors. The dot-dashed line is for models in which the matter power
spectrum is normalized to 8 h$^{-1}$ Mpc scales. The high dependence
of its value and error on $\Omega_m$ causes the 
$\sigma_8$ normalized limits to be different
from those in which power spectra are normalized to COBE measurements.
In Fig.~3, we have shown the limits assuming a survey of 100 $\times$
100 sqr. degrees down to a R band magnitude of 25.
 However, for surveys with different areas,
especially for surveys in near future with small coverage,
the limits can be scaled by the reduction factor
in the observed area (see, Eq.~12). We assume uncorrelated errors in the
weak lensing power spectrum measurement.
For low $\Omega_m$ models ($\la 0.5$) normalized
to COBE, and using our conservative errors,
we can write the 2 $\sigma$ detection 
limit on the neutrino mass as:
\begin{equation}
m_\nu^{2\sigma}
 \sim 5.5 \left(\frac{\Omega_mh^2}{0.1 N}\right)^{0.8}
\left(\frac{10}{\theta_s}\right).
\end{equation}
This 2 $\sigma$ detection limit is comparable to current upper limits
at the 2 $\sigma$ level on the neutrino mass. Using more optimistic
errors decreases this limit by a factor of 2 to 3 depending
on $\Omega_m$, however, to obtain such optimistic errors on
cosmological parameters one require accurate measurements on the
CMB power spectrum such as to the level of MAP satellite.
In making this prediction we have assumed that the weak lensing
power spectrum can be measured to the expected uncertainty predicted
by simple arguments involving errors in ellipticities and
cosmic confusion and that the measurements are uncorrelated. 
Also, in order to obtain a reliable measurement of the
weak lensing power spectrum, one require additional knowledge
on the redshift distribution of sources. Such information is
likely to be adequately obtained with photometric redshift
measurements of color data or by template fitting
techniques that has been developed for multicolor
surveys (e.g., Hogg et al. 1998). The accuracy to which
such measurements can be made should be  adequate, however, if
no multicolor data is available then this may not be possible.
Therefore, it is likely that such a {\it clean} measurement of the weak lensing
power spectrum will not be directly possible in the near future. 
In order to consider
such affects, we increased the expected uncertainties in the power
spectrum by a factor of 2 beyond what is predicted for a survey of
100 sqr. degrees down to a R band magnitude of 25. The expected
neutrino mass limit increases by an amount consistent with
what is expected from the Fisher matrix formalism. Even in such a
scenario with a poorly measured power spectrum, one can still put
interesting limits on the neutrino mass.

A  small area survey such as 10 $\times$ 10
sqr. degrees is likely to be feasible
in the near future with upcoming observations from wide
field CCD cameras. There are several such instruments currently
either in the design or manufacturing stages: MEGACAM\footnote{
http://cdsweb.u-strasbg.fr:2001/projects/megacam/} which
will make observations from the Canada France Hawaii Telescope 
(CFHT; Boulade et al. 1998), VLT-Survey-Telescope\footnote{
http://oacosf.na.astro.it/vst/} (VST). Other than these surveys,
which are likely to first produce deep weak lensing  surveys over
small areas, two wide-field shallow surveys are currently ongoing 
at optical (SDSS; Stebbins et al. 1997) and radio (FIRST;
Kamionkowski et al. 1997),  however, it is still unclear as to what accuracy
these imaging data can be used for weak lensing studies.
Still, assuming that SDSS can in fact make weak lensing measurements
down to a R band magnitude of 22, we find that given its wide field
coverage, it can also be used to detect neutrinos down to a mass limit
of $\sim$ 3 eV at the 2$\sigma$ level,  
or to put interesting limits at the same mass threshold.
For an ultimate survey of $100 \times 100$ deg$^2$, weak lensing
allows a detection of neutrinos down to a mass of $\sim$ 0.5 eV
when $\Omega_m \sim 0.3$ and $h \sim 0.65$. With expected errors
from CMB satellites, this limit can be lowered by a factor of 3 to 4
allowing a possibility for weak lensing surveys to probe neutrinos
with mass lower than 0.1 eV.  These conclusions, generally, are
consistent with what was found by Hu \& Tegmark (1999); minor
differences are likely to arise from the fact that the present study
and Hu \& Tegmark (1999) used different fitting functions to describe
the non linear evolution of the potential power spectrum and
that fudicial cosmological models may be different. 
We note here that using MAP 
or PLANCK data with galaxy 
redshift surveys such as from SDSS, and no weak lensing
measurements, only allow the determination of neutrino mass
to a limit of $\sim$ 1 eV and 0.3 eV respectively (Hu et al. 1997).

\begin{figure}
\centerline{\psfig{file=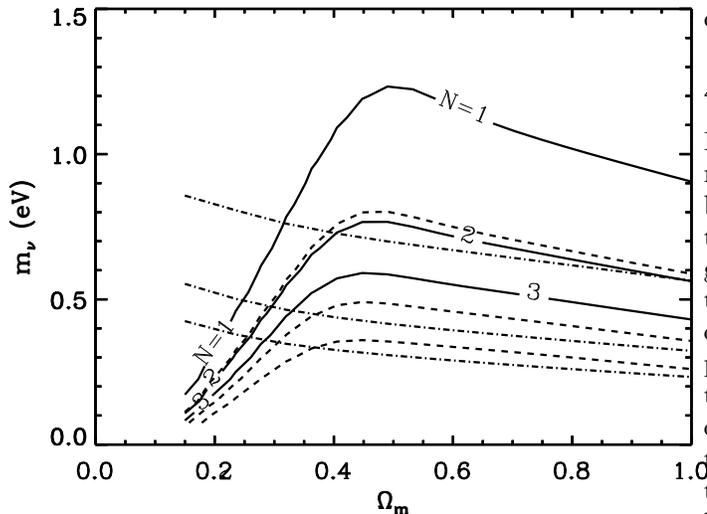,width=4.1in,angle=90}}
\caption{Expected 2$\sigma$ detection limit for 100 $\times$ 100 deg.$^2$ weak
lensing survey down to a magnitude limit of 25, assuming
a spatially flat Universe with a scalar tilt of 1. Solid line
represents detection with our conservative errors while
the dashed line represent detection with more optimistic errors.
The dot-dashed line is for models involving normalizations based
on current measurements of $\sigma_8$ and using more optimistic
errors.}
\end{figure}

Returning to much smaller surveys, we have only studied the accuracy to 
which the neutrino mass can be measured. However, in making such
measurements one does not lose information to make other measurements
as well. For example, the conservative errors we assumed on other
cosmological parameters can also be improved by factors of 2 to 3 when
information on these parameters are also derived with weak lensing.
Also, one can abandon the assumption of  a spatially flat
Universe, and determine the value of the cosmological constant
directly from weak lensing data, while also putting a limit on
the neutrino mass. However, if the assumption on a spatially
flat Universe is dropped in order to measure $\Omega_\Lambda$, then
the limit to which neutrino mass can be measured increases by a
factor of $\sim$ 1.5 for surveys of size 100 sqr. degrees. For now,
if one to measure or improve all other cosmological parameters
that can be studied with weak lensing surveys (and listed in
Sect.~2.1), then it is safe to say that neutrinos down to a mass limit of
$\sim$ 8 eV can be measured with weak lensing surveys of size 100
sqr. degrees down to a R band magnitude of 25. Such a possibility will
definitely  be available with upcoming surveys from
MEGACAM. For still smaller surveys, such as 10 sqr. degrees, if one
attempts to make all cosmological measurements, such as
$\Omega_m$ and $\Omega_\Lambda$ interesting limits on neutrino
mass can only be obtained at a mass level greater than 25 eV.
 Since such neutrino masses may be ruled out, it is safe to ignore the
presence of neutrinos when making measurements with much smaller surveys.
Such surveys are likely to be first available with wide-field cameras,
with the coverage increasing afterwards.

\section{Discussion \& Summary}

Here, we have considered the possibility for a neutrino mass measurement
using weak gravitational lensing of background sources due
to foreground large scale structure. For survey of size 100
deg$^2$, neutrinos with masses greater than $\sim$ 5.5 eV could easily be
detected. This detection limit is comparable to the current
cosmological limits on neutrino mass, such as from the Ly$\alpha$ forest.
 When compared to various ongoing experiments
to detect neutrinos, the advantage of weak lensing 
is that one can directly obtain a measure of mass rather than mass 
difference between two neutrino species.
For typical  surveys of size $\sim$ ten square degrees, ignoring the
presence of neutrinos can lead to biased estimates for cosmological 
parameters, e.g., cosmological mass density can be underestimated
by a factor as high as $\sim$ 15\% if neutrinos
with mass 5 eV are in fact present.  However, if such weak lensing
surveys  are solely used for the  derivation of parameters such
as cosmological mass density,  than the accuracy to which such 
derivations can be made is less than the bias produced by 
neutrinos. Therefore, for small area
surveys, the presence of neutrinos can be
safely  ignored (assuming that their masses is less than $\sim$ 5 eV or so).
However, armed with cosmological parameters from other complimentary
techniques, even such small weak lensing surveys allow a strong
possibility to investigate the presence of non-zero mass neutrinos.

\begin{acknowledgements}

We acknowledge useful discussions with Wayne Hu
and Dragan Huterer. Wayne Hu is also thanked for communicating
the fitting code to evaluate the MDM transfer function.
We also thank an anonymous referee for comments which led to
several improvements in the presentation.

\end{acknowledgements}

\end{document}